\documentclass[12pt]{article}
\pdfoutput=1 

\usepackage{jheppub} 

\usepackage[T1]{fontenc} 
\usepackage{soul}  

\usepackage{etoolbox}
    \makeatletter
    \patchcmd{\maketitle}{\@fpheader}{\vspace{1cm}}{}{}
    \makeatother

\def\obs{\mathcal{O}}

\title{\boldmath 
Gravity from micro-equilibration
}

\author[a]{Veronika E.\ Hubeny}
\author[b]{and Massimiliano Rota}
\affiliation[a]{Center for Quantum Mathematics and Physics (QMAP)\\
Department of Physics, University of California, Davis, CA 95616 USA}
\affiliation[b]{Department of Physics, University of California, Santa Barbara, CA 93106, USA
\\
\\
}

\emailAdd{veronika@physics.ucdavis.edu}
\emailAdd{mrota@physics.ucsb.edu}

\abstract{
We suggest a mechanism for the emergence of classical dynamical spacetime from an underlying quantum gravitational system. This is an example of a more general process, which we name micro-equilibration, and which can be thought of as local thermalization of entanglement.  Applied in the context of the AdS/CFT correspondence, we propose that this dynamical process underlies generic evolution towards CFT states whose gravitational dual describes smooth bulk geometries.  Hence contrary to common expectation, such `geometric' CFT states are in fact typical in this sense.  Correspondingly they can be characterized by a specific universal entanglement structure resulting from micro-equilibration of a generic quantum state.
\\
\\
\\
\\
\\
\\
\\
\\
\\
\\ 
\small{{\it Essay written for the Gravity Research Foundation 2018 Awards for Essays on Gravitation}}
\newline
Submitted on: March 29, 2018  
}
\notoc 

\begin{document} 
\maketitle
\flushbottom

\section{Introduction}
\label{s:}

The desire to elucidate the emergence of a classical spacetime from a more fundamental quantum gravitational description has motivated numerous exhilarating research directions over the last few decades.  A particularly fruitful avenue was paved by the holographic AdS/CFT correspondence \cite{Maldacena:aa}, which recasts quantum gravity (string theory) in the bulk of asymptotically Anti de Sitter  (AdS) spacetime in terms of a lower-dimensional conformal quantum field theory (CFT, naturally conceptualized as living on the boundary of AdS).  Successful implementation of this holographic duality to explain how the bulk gravitational degrees of freedom emerge from the boundary CFT requires detailed understanding of the map between the two sides. One of the key questions in this context is,  {\it Given a CFT with holographic dual, what characterizes the class of CFT states which represent classical bulk geometries?}  Recent developments \cite{Raamsdonk:2010aa,Maldacena:2013aa} suggest that entanglement plays a crucial role, so a more 
refined 
question
  concerns the entanglement structure of such `geometric' CFT states.
  Answering this question would bring us closer to understanding interesting departures from classicality such as those occurring near curvature singularities.

A natural expectation, often implicit in these discussions, is that geometric CFT states are extremely special.  After all, generic superpositions of states corresponding to distinct bulk geometries cannot themselves be associated with a single classical bulk spacetime.  However, one could have used the same reasoning to argue for genericity of superpositions in any quantum system -- in particular, in our world -- which is naively at odds with the success of classical physics, and motivates the foundational question regarding the emergence of classicality from the quantum:  {\it Why don't we see macroscopic superpositions, such as Schr\"odinger cats, all around us?}  

In this essay we propose that these two questions are in fact closely related in the holographic context:  classical bulk geometry is not a highly fine-tuned class of special states in the CFT, but rather something that emerges very naturally, by exactly the same mechanism responsible for emergence of classicality from our quantum world.  

In what follows, we propose a conceptual framework for this mechanism, which we will dub `micro-equilibration'.  To this end, we will first characterize classicality, then explain the dynamical mechanism whereby it is attained, and finally return to the implications for holography and more general spacetime emergence.

\section{Classicality}
\label{s:}

The notion of classicality is so familiar that we are seldom faced with the task of defining it.  However, in order to explain how classical behavior can be emergent, a careful definition is needed. There are two apparently distinct explanations for why our world looks classical: 
\begin{enumerate}
\item
The observables we have access to are not sensitive enough to discern the ever-present quantum superpositions.
\item
There is a dynamical mechanism whereby such superpositions become suppressed. 
\end{enumerate}
%

Zurek \cite{Zurek:aa} proposed an explanation which has aspects of both:  a system decoheres due to interactions with its environment, which leads to an environment-induced superselection (or {\it einselection} for short) into a set of `pointer states' that are stable under further interactions.  However, only certain (classical) information about a quantum system can be redundantly proliferated to many parts of its environment, thereby becoming accessible and objective.\footnote{
The idea that  environment actively selects such information is known as 
 `quantum Darwinism' \cite{Zurek:aa}, with Darwinistic fitness corresponding to the amount of redundancy, providing a measure of classicality. 
}

Here we espouse a more direct notion of classicality which explicitly
evokes both states and observables. Since the essence of quantumness is rooted in the possibility of superpositions, we expect that classicality should manifest itself as absence of superpositions; however such condition by itself is basis-dependent and therefore ill-defined.  To evoke only physical quantities, we instead utilize a closely-related feature of classicality: the same measurement performed on the same state should yield the same outcome.

More precisely, suppose that a (restricted) set $\obs_{cl}$
of observables has been specified; for example one can imagine that $\obs_{cl}$ captures the limited capabilities of an experimentalist performing measurements in a laboratory. If multiple copies of a system are prepared in the same state, we will say that the observations `look classical' if for any observable in the set $\obs_{cl}$, the same outcome is attained for each copy, at least to a good approximation.

Note that according to this definition, classicality can be considered an attribute of a state only once a set of observables has been specified. 
While sensible for the experimentalist, a more fundamental theory should also explain how the set $\obs_{cl}$ itself can be identified. As proposed in the next section, a natural definition of $\obs_{cl}$ emerges once we  include the effects of 
the dynamics.

\section{Micro-equilibration}
\label{s:}

To motivate the proposed mechanism,
let us first consider an analogy with classical statistical mechanics. Initial dynamics often drives systems with many degrees of freedom towards some attractor set, forming a lower-dimensional subspace of the phase space.  After a sufficiently long time, interacting systems typically thermalize, since large spatial variations tend to get  washed out by the interactions. Generically, this can happen in multiple stages on different time scales, through a sequence of distinct equilibrium states 
before full thermalization is attained. 

A similar picture applies to quantum mechanical systems as well, even at the microscopic level. 
For a given set of observables $\obs$ we can introduce, following \cite{Short:2011aa}, a measure of distinguishability\footnote{ $\mathcal{D}_{\obs}(\rho,\sigma)=\max_{M\in\obs}\mathcal{D}_{M}(\rho,\sigma)$ where $\mathcal{D}_{M}(\rho,\sigma)=\sum_r|\text{Tr}(M_r\rho)-\text{Tr}(M_r\sigma)|/2$ and $M_r$ are positive operators corresponding to all possible outcomes of $M$ satisfying $\sum_r M_r=I$. 
This measure reduces to the trace norm when $\obs$ is the set of all possible observables. }
$\mathcal{D}_{\obs}(\rho,\sigma)$ 
 of two quantum states (or generally density matrices) $\rho$ and $\sigma$. Calling $\omega_T$ the time-averaged state for the time interval $\left[0,T\right]$, the system is said to have equilibrated from the point of view of the observables in $\obs$, if for most times $t \in [0,T]$, $\mathcal{D}_{\obs}(\rho (t),\omega_T)\ll 1$.
 It was proved in \cite{Short:2011aa} (under certain mild assumptions) that generic systems do equilibrate in this sense.
The generalization to a sequence of distinguishable equilibrium steps $\{\omega_{T_i}\}$ is straightforward.

Our heuristic picture for how such equilibration mechanism could operate at the microscopic level
is the following.
Monogamy of quantum entanglement suggests that local interactions entangle nearby degrees of freedom while simultaneously suppressing long-range entanglement. Therefore we expect that the resulting entanglement structure itself will locally equilibrate, on microscopic time scales, catalyzing decoherence. 

Since this process concerns the most fine-grained details of the quantum state, we dub it {\it micro-equilibration},\footnote{
In quantum information language, micro-equilibration can then be thought of as suppression of quantum discord defined in \cite{Zwolak:2013aa}.} and we 
denote the corresponding set of observables $\obs_{meq}$.
Within $\obs_{meq}$ we then pick all the observables for which the above classicality condition is satisfied; this identifies the desired set $\obs_{cl}\subseteq\obs_{meq}$ characterizing the quantum to classical transition.

Micro-equilibration also suggests a more intuitive notion of typicality, which resolves the apparent paradox posed in the Introduction:  if (in any given basis) a state picked at random 
from the Hilbert space is generically a superposition, why does the world nevertheless appear classical?  Our answer is that from the point of view of an appropriate class of observables, such a random state would micro-equilibrate to appear classical on incredibly short timescales.
 Hence if we define a ``typical state'' as one which is picked at random not at a single time instant, but rather from many instants of time (over timescales much longer than the micro-equilibration time though still much shorter than those characterizing classical dynamics), then the typical state would indeed reflect our experience of classicality.  According to this prescription, typicality is then quite distinct from randomness.
 
\section{Holography}
\label{s:}

Let us now return to the question posed at the beginning of this essay, regarding the characterization of `geometric CFT states', defined as those representing classical bulk geometries.  Since in the relevant regime the CFT is a strongly interacting system with many degrees of freedom, we would expect micro-equilibration to be operative particularly efficiently, indeed maximally quickly.  The micro-equilibrated phase will have a classical description, 
naturally 
manifested by the bulk dual featuring a  classical geometry.\footnote{
Indeed, this 
suggests the underpinning for why the bulk degrees of freedom repackage the CFT ones so elegantly.
} This means that contrary to the naive expectations indicated earlier, CFT states with  geometrical bulk dual are actually typical in the above sense.

This observation also provides a ready means to answer what entanglement structure characterizes such states:  it is the one given by the micro-equilibrated entanglement structure, obtained by evolving a generic
initial state.
Because of its universality, this motivates why e.g.\ entanglement entropy can satisfy certain interesting relations (such as monogamy of mutual information \cite{Hayden:2011aa}) which need not be satisfied by non-geometric quantum states.

In closing, we remark that
although our discussion was motivated in the holographic context, the indicated behavior in fact transcends holography.  The mechanism suggested above should be operative in any interacting quantum system.  For systems describing gravity,  micro-equilibration should 
 realize
the emergence of a classical dynamical spacetime.  
It has been repeatedly observed \cite{Bekenstein:1973ur,Jacobson:1995ab,Wald:1999xu} that spacetime has thermodynamic properties.  Here we propose that this is rooted in micro-equilibration, which may be thought of as thermalization of entanglement.



\section*{Acknowledgements}

We are delighted to thank Andy Albrecht for collaboration and many fruitful discussions.  We would also like to acknowledge related stimulating discussions with the participants of the KITP {\it Quantum Physics of Information} program, especially Ahmed Almheiri, C\'edric B\'eny, Fernando Brandao, Xi Dong, Juan Maldacena, Don Marolf, Fernando Pastawski, Mukund Rangamani, and Wojciech Zurek.
We thank 
the Yukawa Institute for Theoretical Physics, 
the Perimeter Institute for Theoretical Physics, 
the Kavli Institute For Theoretical Physics,
and the Abdus Salam International Centre for Theoretical Physics
for hospitality during various stages of this project.
VH is supported in part by U.S. Department of Energy grant DE-SC0009999. MR is supported by the Simons Foundation via the ``It from Qubit'' collaboration and by funds from the University of California.
\\
\\
\small{\textit{{
This essay is dedicated to the fond memory of a brilliant physicist, kind friend and inspiring mentor, Joe Polchinski,  ever gracious in sharing his insights and fearless in playful speculation.}}}         

\newpage


\bibliographystyle{jhep}
\bibliography{essayrefs}

\providecommand{\href}[2]{#2}\begingroup\raggedright\begin{thebibliography}{10}

\bibitem{Maldacena:aa}
J.~M. Maldacena, {\it The large n limit of superconformal field theories and
  supergravity},  \href{http://arxiv.org/abs/hep-th/9711200}{{\tt
  arXiv:hep-th/9711200}}.

\bibitem{Raamsdonk:2010aa}
M.~V. Raamsdonk, {\it Building up spacetime with quantum entanglement},
  \href{http://arxiv.org/abs/1005.3035}{{\tt arXiv:1005.3035}}.

\bibitem{Maldacena:2013aa}
J.~Maldacena and L.~Susskind, {\it Cool horizons for entangled black holes},
  \href{http://arxiv.org/abs/1306.0533}{{\tt arXiv:1306.0533}}.

\bibitem{Zurek:aa}
W.~H. Zurek, {\it Decoherence, einselection, and the quantum origins of the
  classical},  \href{http://arxiv.org/abs/quant-ph/0105127}{{\tt
  arXiv:quant-ph/0105127}}.

\bibitem{Short:2011aa}
A.~J. Short and T.~C. Farrelly, {\it Quantum equilibration in finite time},
  \href{http://arxiv.org/abs/1110.5759}{{\tt arXiv:1110.5759}}.

\bibitem{Zwolak:2013aa}
M.~Zwolak and W.~H. Zurek, {\it Complementarity of quantum discord and
  classically accessible information},
  \href{http://arxiv.org/abs/1303.4659}{{\tt arXiv:1303.4659}}.

\bibitem{Hayden:2011aa}
P.~Hayden, M.~Headrick, and A.~Maloney, {\it Holographic mutual information is
  monogamous},  \href{http://arxiv.org/abs/1107.2940}{{\tt arXiv:1107.2940}}.

\bibitem{Bekenstein:1973ur}
J.~D. Bekenstein, {\it {Black holes and entropy}},  {\em Phys. Rev.} {\bf D7}
  (1973) 2333--2346.

\bibitem{Jacobson:1995ab}
T.~Jacobson, {\it {Thermodynamics of space-time: The Einstein equation of
  state}},  {\em Phys. Rev. Lett.} {\bf 75} (1995) 1260--1263,
  [\href{http://arxiv.org/abs/gr-qc/9504004}{{\tt arXiv:gr-qc/9504004}}].

\bibitem{Wald:1999xu}
R.~M. Wald, {\it {Gravitation, thermodynamics, and quantum theory}},  {\em
  Class. Quant. Grav.} {\bf 16} (1999) A177--A190,
  [\href{http://arxiv.org/abs/gr-qc/9901033}{{\tt arXiv:gr-qc/9901033}}].

\end{thebibliography}\endgroup

\end{document}